# Direct and converse magnetoelectric effects in Metglas/LiNbO$_3$/Metglas trilayers

A.A. Timopheev*[1], J.V. Vidal[1], A.L. Kholkin[2] and N.A. Sobolev[1]

[1] *Departamento de Física and I3N, Universidade de Aveiro, 3810-193 Aveiro, Portugal*

[2] *Departamento de Engenharia de Materiais e Cerâmica and CICECO, Universidade de Aveiro, 3810-193 Aveiro, Portugal*

*E-mail: andreyt@ua.pt

Electromechanical and magnetoelectric properties of Metglas/LiNbO$_3$/Metglas trilayers have been studied in the frequency range from 20 Hz to 0.4 MHz. A trilayer of Metglas/PMN-PT/Metglas prepared in the same way was used as a reference. Though PMN-PT has much larger charge piezocoefficients than LiNbO$_3$ (LNO), the direct magnetoelectric voltage coefficient is found to be comparable in both trilayers due to the much lower dielectric permittivity of LNO. The magnitude of the direct magnetoelectric effect in the LNO trilayers is about 0.4 V/cm·Oe in the quasistatic regime and about 90 V/cm·Oe at electromechanical resonance. Calculations show that the magnetoelectric properties can be significantly improved (up to 500 V/cm·Oe) via controlling the cut angle of LNO, choosing the appropriate thickness ratio of the ferroelectric/ferromagnetic layers, and a better bonding between Metglas and LNO. Advantages of using LiNbO$_3$-type ferroelectrics in magnetoelectric composites are discussed.

1. **Introduction**

Magnetoelectric properties of composite multiferroics are being intensively investigated today [1]. Though the underlying physics of the single-phase multiferroics is much deeper, the observed magnitudes of the magnetoelectric (ME) effect as well as typical temperatures of its experimental observation are by far too low for any reasonable practical application. In composite multiferroics [2-4], the required properties of the ferroelectric (FE) and ferromagnetic (FM) phases can be adjusted separately. As the estimated magnitude of the direct ME effect is proportional to the product [5] of the magnetostrictive and piezoelectric properties of the FM and FE phases, carefully designed composites have an extraordinary ME response by orders of magnitude higher than that in single-phase multiferroics [2-4, 6]. The same tendency is observed for the converse ME effect [2-6]. While the latter effect is useful for the electric field control of magnetic properties [7, 8], the direct

ME effect allows building ultrasensitive magnetic field sensors [9-12]. It is worth noting that the spectrum of possible applications of multiferroic composites is obviously much wider, including memory, domain wall motion control, acoustically driven ferromagnetic resonance (FMR), etc. [13-16].

Among different connectivity types of the FM and FE phases in ME composites [2-4], the 2-2 type laminate structures demonstrate the maximum efficiency due to low leakage currents and a strong mechanical coupling between the FE and FM phases. Additional improvements could be achieved by a proper choice of the FE and FM phases' properties. For example, the low saturation field of the FM phase, $H_s$, is usually more important than the high magnetostriction constant, $\lambda_s$. Though the Terfenol-D and related magnetostrictive alloys [17] have two orders of magnitude higher $\lambda_s$ values than different types of metallic glasses [18, 19], the magnetic fields needed to saturate these systems are typically above 1 kOe, which is by two or even three orders of magnitude higher than in abovementioned amorphous alloys. As a result, the magnetoelectric voltage coefficient, $\alpha_E$, in laminates based on metallic glasses can even be higher, which is extremely important for the ultra-low magnetic field sensing applications.

The "L-T" operation mode, in which the FM layer is longitudinally magnetized and the FE layer is perpendicularly polarized, is often used in laminate structures [2-4]. Necessary requirements for the FE phase are a high $d_{31}$ piezocoefficient for the direct ME effect and the condition $d_{31} \neq d_{32}$ for the converse ME effect, in order to achieve an effective generation of the in-plane stress. Though the frequently used $Pb(Zr,Ti)O_3$ (PZT) ceramics does not satisfy these requirements from a general point of view, the design of PZT-based ME composites and multilayers is still ongoing. Remarkable progress has been achieved with a special geometry of electrodes, in the so-called "push-pull" mode [20, 21], allowing the laminate to work effectively in the length extensional resonance. However, the question of miniaturization of such a complicated structure is still open. Another important aspect is nonlinear magnetoelectric effect that can help avoid DC magnetic bias, yet giving high magnetoelectric response at second harmonics [22].

The use of single crystalline ferroelectrics and piezoelectrics [23, 24] poled and cut along desirable crystallographic directions is another way to achieve a strong ME effect. Among different commercially available single crystals, the lead magnesium niobate – lead titanate (PMN-PT) [25] and lead zinc niobate – lead niobate (PZN-PT) demonstrate the highest piezocoefficients, and thus are frequently used in the design of different laminate ME structures [11, 12, 26-30]. The main disadvantages of PMN-PT and PZN-PT are low Curie and depolarization temperatures (~100°C), chemical and electrical instabilities, non-linear behavior, uneasy growth of high-quality crystals and a very high price. Due to these factors, other single crystalline ferroelectrics or just piezoelectrics are being tested at present [24, 31]. Another useful alternative is lithium niobate (LiNbO$_3$, LNO) [32], an uniaxial ferroelectric with a very high Curie temperature (1500 K). Among other attractive features, one should note its relatively low price, high chemical, thermal and mechanical stability, availability of big crystals of sufficiently high quality and, last but not least, its lead-free composition. LNO and the related material LiTaO$_3$ are widely used in surface and bulk acoustic wave devices, optical modulators and waveguides, filters, transducers, acoustic microscopes, etc. [33-35]. Even magnetically tuned surface acoustic wave devices were designed and tested in doped LiNbO$_3$ [15, 36-39]. However, as a functional layer in ME laminate structures, it has been considered and/or implemented only in a few studies [40-42]. Calculations of the ME response for a Terfenol/LNO/Terfenol structure were performed in Ref. [40]. An experimental study of such a structure was done in Ref. [42]. A noticeable tunability of the magnetic properties of a polycrystalline nickel film deposited atop a single-crystalline LNO substrate has been shown experimentally in Ref. [41]. Though the piezocoefficients of LNO are much lower than those of PMN-PT and PZN-PT, the dielectric constant is also much lower, which should yield a notable magnetoelectric voltage coefficient in direct ME effect measurements. Even in the weakly piezoelectric quartz [24] the magnetoelectric coefficient has been found to be very high (~175 V/(Oe·cm)) at the electromechanical resonance. Optimization of the LNO-based structures is believed to yield comparable or even larger magnetoelectric coefficients. Multifunctionality of LNO may bring about new opportunities for the coupling between optical, magnetic and electric signals, as well as for corresponding sensing capabilities.

In this work, we present a study of the direct and converse magnetoelectric effects for a trilayer structure of Metglas/LiNbO$_3$/Metglas. As the thickness of the FM layer was not optimized and, besides, the FM/FE bonding method is believed to be far from optimum, another trilayer of Metglas/PMN-PT(011)/Metglas was prepared under the same conditions just to serve as a reference ME sample for comparison.

## 2. Experimental methods and samples

Several "2-2" connectivity type trilayer ME composites were fabricated by bonding foils of an amorphous ferromagnetic alloy to opposite sides of the square plates (10×10×0.5 mm$^3$) of ferroelectric single crystals of LiNbO$_3$ and PMN-PT (PT content = 0.31). Commercially available 29 μm thick sheets of the metglas (2826MB Metglas® by Hitachi Metals Europe GmbH) were attached to the opposite surfaces of LiNbO$_3$ single crystals (Roditi International Corporation Ltd.)

and PMN-PT (H.C. Materials Corp.) using a commercial cyanoacrylate-based glue. The LiNbO$_3$ crystal has a trigonal symmetry, and the polarization vector lies along the c-axis. LiNbO$_3$ samples were poled by the supplier immediately after growth. After the poling procedure, the LNO crystals remain in the single-domain state, and repolarization/depolarization does not occur at room temperature (RT). The main physical properties of LNO can be found in Ref. [32]. The data presented there are valid for Z-cut substrates only. In our study we used Y-cut and 41°Y-cut crystals, whose physical properties (with respect to the sample geometry) are poorly represented in the literature. Thus, the piezocoefficients, dielectric permittivity and elastic compliances of the used LNO substrates will be calculated below. (011)-cut PMN-PT plates were chosen as a reference. The prepared trilayer also operates in the L-T ME mode, i.e. with the transverse polarized piezoelectric phase (out-of-plane geometry) and longitudinal magnetized magnetostrictive phase (in-plane geometry). The PMN-PT crystals used in these measurements were poled beforehand along their thickness (i.e., parallel to the <011> direction), so that the *mm*2 multidomain symmetry was engineered [25] with the effective in-plane extensional piezocoefficients d$_{31}$ = −1700 pC/N and d$_{32}$ = 850 pC/N.

To study the electromechanical and ME properties of the trilayers, impedance spectroscopy, as well as direct and converse ME effect measurements were carried out. The impedance measurements were performed in the frequency range from 0 – 10 MHz. A reference resistor was connected in series with a trilayer sample, and a frequency sweep with a constant voltage was performed by a function generator. The voltage amplitudes and phases on the reference resistor and sample were detected, and then, using a simple equivalent circuit model, the active and reactive parts of the sample impedance were obtained. This technique was used mainly to determine the quasi-static dielectric constant of the trilayers and to find the frequencies of the electromechanical resonances.

Figure 1 schematically shows the experimental setup built for the measurements of the direct and converse ME effects. To measure the direct ME effect, the trilayers were placed in the center of a Helmholtz coil generating a small AC magnetic field with the amplitude $\delta H$. The AC ($\delta H$) and DC ($H$) magnetic fields, created by the Helmholtz coil and the electromagnet, respectively, are collinear and aligned along the *x*-direction. The DC field $H$ was stabilized by a Hall-sensor-based regulating feedback loop. The Helmholtz coil was driven by a power amplifier working in the current stabilization mode. The amplitude and frequency of the AC current was set by a function generator (SRS®, model DS345). The AC current was continuously measured by an ammeter (Agilent®, model 34401A), and its value was used to calibrate the generated magnetic AC field amplitude $\delta H$ throughout a large frequency range (up to 500 kHz). A gaussmeter (DSP®, model 475) was used to measure the exact value of the bias field produced by the electromagnet. The transverse voltage $\delta V$ induced across the sample by the applied in-plane AC magnetic field $\delta H$ was measured with a lock-in amplifier (Zurich Insruments®, model HF2LI). The measuring system as a whole is synchronized by a TTL output signal from the function generator and is driven by a custom-made data acquisition software.

The samples were magnetized in the *x*-y plane, while the induced AC voltage was measured along the *z*-direction. The RT measurements of the direct ME voltage coefficient $\alpha_{E3i} = \delta V/(\delta H \cdot t)$, (with $i = 1,2$; $t$ being the thickness of the piezoelectric crystal) were performed for the trilayers as a function of the bias field *H*. The modulation amplitude of the AC field was fixed at $\delta H = 1$ Oe and its frequency was $f = 5$ kHz. The transverse ME voltage coefficients $\alpha_{E31}$ and $\alpha_{E32}$ were measured in two orthogonal in-plane orientations: $\alpha_{E31}$ was measured with *H* oriented along the <100> crystallographic direction (i.e. <100> ∥ *x* axis), while for the $\alpha_{E32}$ coefficient measurements the sample was rotated in-plane counterclockwise by 90º (i.e. <100> ∥ *y* axis). Beside the magnetic field dependences ($\alpha_{E31}(H)$ and $\alpha_{E32}(H)$), the measurements of $\alpha_{E31}$ and $\alpha_{E32}$ were conducted also as a function of the AC magnetic field frequency, *f*, in the range from 20 Hz – 400 kHz (with a fixed bias field $H = 25 - 30$ Oe and the AC field amplitude $\delta H = 1$ Oe). In both type of measurements a Faraday induction voltage has been detected additionally to the direct ME response. It results from the use of an AC excitation magnetic field and is dependent on the concrete geometry of the experiment. It was established that the induction amplitude linearly increased with the frequency with a slope of 108 mV/(MHz·Oe) in all our measurements. For the most experimental data presented here, this parasitic effect is negligibly small; otherwise it was subtracted from the data during the post-processing procedure (in the latter case it will be noted in the text).

Measurements of the converse ME effect were performed by applying an AC voltage from the function generator to the sample and using the Helmholtz coil to detect the magnetic flux variation due to the changing magnetic permeability of the sample. The ME voltage amplitude and phase as a function of either the driving frequency at a constant DC magnetic field or the swept magnetic field at a constant excitation frequency were detected by the lock-in. The resulting ME effect was represented as a ratio of the AC voltage detected on the Helmoltz coils to the voltage applied to the sample. The accessible frequency range for this type of measurements spans from a few Hz up to several tens of MHz.

3. **Experimental results and discussion**

The prepared trilayers were first measured using the impedance spectroscopy setup in order to find and to identify the electromechanical resonance modes. The results are shown in Figure 2. The sound velocity for the longitudinal waves in LiNbO$_3$ propagating along the principal axes is: for the Z-direction, 7.271×10$^5$ cm/s; for the Y-direction, 6.549×10$^5$ cm/s; and for the X-direction, 6.580×10$^5$ cm/s [32, 43, 44]. Applying these values to the case of a 10×10×0.5 mm$^3$ Y-cut LNO crystal, the thickness extensional mode should appear at 6.58 MHz. The closest observed peak is centered at a slightly lower frequency of 6.34 MHz, which may be explained as a consequence of the clamping effect of the Metglas foils. The calculated length extensional resonances along the X and Z directions (327.5 kHz and 363.5 kHz, respectively) can be attributed to the peaks observed at 314 kHz and 355 kHz. The two remaining peaks centered at 3.65 MHz and 281 kHz are too far from any possible longitudinal acoustic wave resonances and evidently should be attributed to shear modes propagating through the thickness and length, respectively. For the case of the 41ºY-cut LNO crystal, the shear modes should be suppressed [43, 44], and the two observed peaks must reflect the thickness and length extensional modes for longitudinal waves. The calculated dielectric constants, $\varepsilon_{33}$ (along the substrate thickness), for quasi-static off-resonance conditions ($f$ = 150 kHz) are 69 and 45 for the Y-cut and 41ºY-cut, respectively.

For the case of the PMN-PT-based trilayer, two low-frequency modes (67 kHz and 111 kHz) and one high-frequency mode (4.87 MHz) are observed. According to Ref. [45], the sound velocity for the longitudinal mode propagating along the <011> direction is 4.727 cm/s, which yields for a thickness extensional resonance a frequency of about 4.73 MHz that is close to the experimentally observed value. We were not able to exactly identify the types of the low-frequency peaks, however, there is no doubt that both of them represent certain types of shear modes. The dielectric constant $\varepsilon_{33}$ (along the substrate thickness) measured at $f$ = 1 kHz is 4440.

Measurements of the direct ME effect were conducted at RT at $f$ = 5 kHz, sufficiently far from the observed electromechanical resonances, i.e. in a quasi-static regime. Figure 3 demonstrates the experimental results. There are two points that should be noted:

(i) Soft magnetic properties of the metglas film provide the maximum ME effect in magnetic fields as low as 25 Oe. The maximum corresponds to the field range where the magnetization grows by the rotation of magnetization vectors inside the magnetic domains and by the domain structure rebuilding. The saturation of the ferromagnetic film occurs at $H \approx$ 50 Oe, after which further growth of the external field does not lead to any substantial increase of the stresses in the film, so that the dynamical part of the ME effect vanishes.

(ii) The samples demonstrate anisotropy of the in-plane ME properties. The trilayer prepared from the Y-cut LiNbO$_3$ crystal exhibits a difference between $\alpha_{E31}$ (***H*** || <100>) and $\alpha_{E32}$ (***H*** || <001>) by more than an order of magnitude. The obtained values are: $\alpha_{E31}$ = +0.46 V/(cm·Oe) and $\alpha_{E32}$ = –0.024 V/(cm·Oe). For the <001> direction, the ME response is comparable to the parasitic Faraday induction voltage, so that the corresponding correction has been applied to the measured data. In contrast, for the 41°Y-cut LNO crystal an isotropic in-plane behavior has been observed: for both in-plane directions, $\alpha_{E31}$ and $\alpha_{E32}$ are approximately the same: $\alpha_{E31} \approx \alpha_{E32} \approx$ +0.42 V/(cm·Oe). Such noticeable changes are mainly linked to the anisotropy of the piezoconstants of LNO. Thus, different crystal cuts yield strikingly different magnetoelectric properties for the trilayers. Calculations of piezoconstants and magnetoelectric coefficients as a function of the crystal cut angle for LNO-based trilayers will be given below. In the case of the PMN-PT trilayer, the $d_{31}$ piezoconstant is twice as high as $d_{32}$ and has a different sign. As a result, the maximum of $\alpha_{E31}$ (***H*** || <100>) corresponds to +1.15 V/(cm·Oe), while the maximal amplitude for $\alpha_{E32}$ (***H*** || <011>) is –0.41 V/(cm·Oe). The obtained values are relatively small, and the main reasons for this are a too high piezoelectric/ferromagnetic thickness ratio and non-optimal bonding between the constituent phases. Following the procedure described in Ref. [40] and using the Metglas and PMN-PT material parameters obtained from the suppliers and from independent sources [6, 45-49], we calculated the maximal values of $\alpha_{E31}$ and $\alpha_{E32}$ for our specific samples. The obtained values were: $\alpha_{E31}$ = 23.2 V/(cm·Oe) and $\alpha_{E32}$ = –7.2 V/(cm·Oe). These values are more than one order of magnitude larger than the experimental ones. This may be due to the weak interfacial bonding in our case, and the measured coefficients may significantly increase if a stronger and thinner adhesive layer can be used. On the other hand, apparent stress relaxation occurring along the thickness of the thick piezoelectric layer is not taken into account in the model [40]. Nevertheless, as the main role of this sample is to serve as a reference and as both factors discussed above should result in the similar reduction of the ME effect in the PMN-PT- and LNO-based trilayers, further discussion will be focused on the measured dependencies rather than on the values themselves.

To achieve a deeper understanding of the observed effects, one needs to calculate the effective piezoconstants for these two LNO substrates. Using the data for the Z-cut crystal [43] we applied a standard method [40,43,44] for the determination of the piezoelectric and other properties in the rotated (i.e. cut) crystal. The obtained physical quantities were put into the model presented in Ref. [40] to calculate the magnetoelectric response of the trilayers. Figure 4a shows the results of the calculations of the $\alpha_{E31}$ and $\alpha_{E32}$ coefficients as a function of the crystal cut angle. As it can be seen, there is a full qualitative agreement with the experiment. For the 41°Y cut crystal, the $\alpha_{E31}$ and $\alpha_{E32}$ coefficients have approximately the same magnitude due to the close values of the recalculated $d_{31}$ and $d_{32}$ piezocoefficients ( −16.5 pC/N and −17.5 pC/N, respectively). For the Y-cut crystal we have obtained $d_{31}$ = −20.8 pC/N and $d_{32}$ = 0. Consequently, the observed low magnitude of the $\alpha_{E32}$ coefficient is only due to the transversal part of the magnetostriction acting on the $d_{31}$ component.

From a general point of view, the use of lithium niobate with different cuts allows one to choose desired anisotropic properties for a magnetoelectric system design. Independently of the chosen cut, the polarization vector will be always aligned along the crystallographic *c*-axis, and a repolarization during the operation is practically impossible. The most interesting cut directions from the viewpoint of magnetic field sensors are: 1) 2.1°Y-cut, where $\alpha_{E31}$ = 11.5 V/(cm·Oe) and $\alpha_{E32}$ = 0 − anisotropic unipolar regime; 2) 44.9°Y-cut, where $\alpha_{E31}$ = $\alpha_{E32}$ = 11.3 V/(cm·Oe) − isotropic regime; 3) 162.1°Y-cut, where $\alpha_{E31}$ = −13 V/(cm·Oe) and $\alpha_{E32}$ = 13 V/(cm·Oe), − anisotropic bipolar regime. The first and last cuts could be useful in vector magnetic field sensors. Also the 129°Y-cut is of great importance as it has the maximal ME voltage coefficient $\alpha_{E32}$ = 27 V/(cm·Oe).

Figure 4b demonstrates the calculated ME voltage coefficients as a function of the ferroelectric/ferromagnetic relative thickness ratio using the same formalism as in Ref. [40]. The calculation was done for the most interesting LNO crystal cuts. The smaller is the thickness of the ferroelectric layer, the more strain is transferred to it from the ferromagnet. However, even if these calculations are absolutely correct quantitatively, the detected voltage is proportional to the ferroelectric layer thickness. Thus, the thinner the layer is, the lower is the output signal. Taking into account that any measuring circuit has its own input noise, the maximal signal-to-noise ratio will be observed at a finite ferroelectric film thickness.

Another important conclusion is that the magnitudes of the observed ME voltage coefficients measured on the LNO trilayers are only three times lower than those for the PMN-PT trilayer. The maximal off-resonance ME voltage coefficient for the PMN-PT trilayer is 1.16 V/(cm·Oe), while it is 0.47 V/(cm·Oe) and 0.421 V/(cm·Oe) for the LNO trilayers produced from Y-cut and 41°Y-cut crystals, respectively. Comparable magnitudes are mainly explained by the fact that the dielectric constant of PMN-PT is more than 60 times higher. Thus, the same amount of generated charge will give a 60 times lower voltage, so that the effectiveness of the huge PMN-PT piezocoefficients is fully masked by the much higher dielectric permittivity.

However, this concurrent influence on the trilayers' performance observed for the direct ME effect is not anymore favorable for the case of the converse ME effect measurements. In this regime, the huge difference in piezoconstants between PMN-PT and $LiNbO_3$ becomes crucial, and the PMN-PT trilayer shows a much stronger converse ME response. Figure 5 shows the converse magnetoelectric measurements represented as a ratio of the voltage generated on the sensing coil (due to the Faraday induction) to the voltage applied to the trilayer. The effect is linked to changes of the effective permeability of the Metglas film under the strain transferred from the piezoelectric substrate. Though this type of measurements is not calibrated yet to the traditionally used (G·cm)/V units, it is still possible to conduct a comparative study, as all our samples are of the same shape, size, and are measured under the same conditions. The dependences of the converse ME effect versus the external magnetic field (Figure 5) are qualitatively similar to those observed in the direct ME effect measurements (Figure 3). Thus, only the high-response direction for each trilayer will be discussed below. The relative amplitudes of the converse ME effect have been found to be noticeably differing: the 41°Y-cut LNO trilayer demonstrates a two orders of magnitude lower amplitude in comparison with the PMN-PT trilayer. Undoubtedly, for the converse ME effect the difference between the piezocoefficients in the trilayers becomes crucial. However, a noticeable difference is observed even between different LNO crystals. The Y-cut trilayer has a six times larger response, pointing to a stronger electromechanical coupling ($k_{31}$) in it. Naturally, under the electromechanical resonance conditions each trilayer will show several orders of magnitude higher converse ME effect. However, the benefits of using PMN-PT crystal with its superb piezoelectric performance for the converse ME effect in such a structure remain indisputable.

Regarding the direct ME effect measurements, the benefits of PMN-PT look not so convincing. Only three times greater ME voltage coefficient in trilayers hardly justifies an order of magnitude higher price and considerably lower Curie temperature (and associated with this instabilities) of the PMN-PT crystal. It has been found that at the electromechanical resonance, the ME performance of the LNO crystals can be even better (see Figure 6). Surprisingly, the 41°Y-cut LNO-based trilayer has exhibited the maximal ME effect of about 90 V/(cm·Oe), while in the PMN-PT trilayer it was only of 70 V/(cm·Oe). In principle, the same trend has been observed while comparing direct ME effect in ferromagnetic/piezoelectric trilayers (Fig. 7 of Ref. [24]) in different materials based on their d/ε ratio.

It is also worth noting that the electromechanical resonance occurs in a very suitable frequency range. From the point of view of possible sensor applications, this means that the LNO-based magnetic sensor can be built using standard low-cost electronic components. As compared to low-temperature ferroelectrics using LNO-based sensors is of utmost importance for high-temperature applications. And, finally, the observed values of the direct ME effect were measured for a non-optimized thickness ratio of the ferroelectric/ferromagnetic components. Optimization of this ratio, as well as a better bonding, should improve the direct ME response by at least one order of magnitude, with the theoretically calculated limit of 490 V/(cm·Oe) in the quasi-static regime.

Interesting possibilities are expected while combiningthe strong magnetoelectric effect observed in LNO-based composites with the unique electrooptic and photorefractive properties of this material.

## 4. Conclusions

In conclusion, we have conducted a study of the electromechanical and magnetoelectric properties of 2-2 type metglass/LiNbO$_3$/metglas trilayers using single-crystalline substrates of different cuts. Metglas foils were bonded to the 10×10×0.5 mm$^3$ ferroelectric substrates using a commercial cyanoacrylate-based glue.

Due to the simplified preparation method and non-optimal ferroelectric/ferromagnetic thickness ratio, the observed ME effect was rather modest. To avoid this problem, an additional Metgalss/PMN-PT/Metglas sample was prepared under the identical conditions and used as a reference sample in this study. Though the trilayer based on the highly piezoelectric PMN-PT has exhibited several orders of magnitude stronger converse ME effect, the measurements of the direct ME effect have shown comparable magnitudes for both types of trilayers, mainly due to the much lower dielectric constant of LNO crystal. At electromechanical resonance, the highest direct ME voltage coefficient of about 90 V/(cm·Oe) has been obtained for the 41°Y-cut LNO trilayer, being ~30% greater than that for the reference PMN-PT trilayer. According to the performed calculations, an optimized trilayer would theoretically yield a magnitude of the direct ME effect of up to 490 V/(cm Oe) in the quasi-static regime. The use of different cuts will in principle allow obtaining the desired in-plane anisotropy of the magnetoelectric response, which could be very useful in vector magnetic field sensors.

Finally, the lithium niobate-based magnetoelectric 2-2 composites, featuring an excellent stability of the chemical and physical properties at high temperatures and a relatively low price, have a great potential to be used in the magnetic field sensing applications in a wide temperature range.

**Acknowledgments**

This work was partially supported by the FCT of Portugal through the projects and grants PEst-C/CTM/LA0025/2011, RECI/FIS-NAN/0183/2012, SFRH/BPD/74086/2010 and SFRH/BD/89097/2012 and European project "Mold-Nanonet".

**References**

1.  G. Srinivasan, Ann. Rev. Mat. Res. **40**, 153 (2010).
2.  J. Ma, J. M. Hu, Z. Li, and C. W. Nan, Adv. Mater. **23**, 1062 (2011).
3.  C.-W. Nan, M. I. Bichurin, S. Dong, D. Viehland, and G. Srinivasan, J. Appl. Phys. **103**, 031101 (2008).
4.  M. Fiebig, J. Phys. D: Appl. Phys. **38**, R123 (2005).
5.  C.-W. Nan, Phys. Rev. B **50**, 6082 (1994).
6.  S. Dong, J. Zhai, J. Li, and D. Viehland, Appl. Phys. Lett. **89**, 252904 (2006).
7.  N. X. Sun and G. Srinivasan, Spin **02**, 1240004 (2012).
8.  C. A. Vaz, J. Phys.: Cond. Matt. **24**, 333201 (2012).
9.  Y. Chen, et al., Appl. Phys. Lett. **99**, 042505 (2011).
10. J. Gao, Z. Wang, Y. Shen, M. Li, Y. Wang, P. Finkel, J. Li, and D. Viehland, Mater. Lett. **82**, 178 (2012).
11. J. Gao, Y. Wang, M. Li, Y. Shen, J. Li, and D. Viehland, Mater. Lett. **85**, 84 (2012).
12. C.-S. Park, K.-H. Cho, M. A. Arat, J. Evey, and S. Priya, J. Appl. Phys. **107**, 094109 (2010).
13. N. A. Pertsev and H. Kohlstedt, Nanotechnology **21**, 475202 (2010).
14. J.-M. Hu, Z. Li, J. Wang, J. Ma, Y. H. Lin, and C. W. Nan, J. Appl. Phys. **108**, 043909 (2010).
15. L. Dreher, M. Weiler, M. Pernpeintner, H. Huebl, R. Gross, M. Brandt, and S. Goennenwein, Phys. Rev. B **86**, 134415 (2012).
16. T. H. Lahtinen, K. J. Franke, and S. van Dijken, Sci. Rep. **2**, 258 (2012).
17. J. Liu, C. Jiang, and H. Xu, Science China Technological Sciences **55**, 1319 (2012).
18. E. D. Delacheisserie, J. Magn. Magn. Mater. **25**, 251 (1982).
19. R. Hasegawa, J. Optoelectron. Adv. Mater. **6**, 503 (2004).
20. M. Li, Y. Wang, D. Hasanyan, J. Li, and D. Viehland, Appl. Phys. Lett. **100**, 132904 (2012).
21. M. Li, D. Hasanyan, Y. Wang, J. Gao, J. Li, and D. Viehland, J. Phys. D: Appl. Phys. **45**, 355002 (2012).
22. L. Y. Fetisov, Y. K. Fetisov, G. Sreenivasulu, and G. Srinivasan, J. Appl. Phys. **113**, 116101 (2013).
23. H. Greve, E. Woltermann, H.-J. Quenzer, B. Wagner, and E. Quandt, Appl. Phys. Lett. **96**, 182501 (2010).
24. G. Sreenivasulu, V. M. Petrov, L. Y. Fetisov, Y. K. Fetisov, and G. Srinivasan, Phys. Rev. B **86**, 214405 (2012).
25. S.-E. Park and T. R. Shrout, J. Appl. Phys. **82**, 1804 (1997).
26. J. Gao, L. Shen, Y. Wang, D. Gray, J. Li, and D. Viehland, J. Appl. Phys. **109**, 074507 (2011).
27. Y. Wang, D. Gray, J. Gao, D. Berry, M. Li, J. Li, D. Viehland, and H. Luo, J. Alloy Comp. **519**, 1 (2012).
28. Y. Chen, T. Fitchorov, Z. Cai, K. S. Ziemer, C. Vittoria, and V. G. Harris, J. Phys. D: Appl. Phys. **43**, 155001 (2010).
29. V. L. Mathe, A. D. Sheikh, and G. Srinivasan, J. Magn. Magn. Mater. **324**, 695 (2012).
30. M. Liu, et al., Adv. Func. Mater. **19**, 1826 (2009).
31. G. Sreenivasulu, L. Y. Fetisov, Y. K. Fetisov, and G. Srinivasan, Appl. Phys. Lett. **100** (2012).
32. R. S. Weis and T. K. Gaylord, Appl. Phys. A - Mater. **37**, 191 (1985).
33. M. Yachi and M. Ono, *IEEE Ultrasonics Symposium, 1995. Proceedings.*, **2**, 1003 (1995).
34. W. P. Robbins and E. M. Simpson, IEEE Trans. Son. Ultrason. **26**, 178 (1979).
35. G. Matsunami, A. Kawamata, H. Hosaka, and T. Morita, Sensor Actuat. A-Phys. **144**, 337 (2008).

36  V. Koeninger, Y. Matsumura, H. H. Uchida, and H. Uchida, J. Alloy Compd. **212**, 581 (1994).
37  N. Obata, T. Kawahata, R. Suzuki, K. Nishimura, H. Uchida, and M. Inoue, Phys. Stat. Sol. (a) **201**, 1973 (2004).
38  W. P. Robbins and A. Young, IEEE Trans. Son. Ultrason. **32**, 423 (1985).
39  K. H. Shin and M. Inoue, J. Korean Phys. Soc. **53**, 1706 (2008).
40  H.-Y. Kuo, A. Slinger, and K. Bhattacharya, Smart Mater. Struc. **19**, 125010 (2010).
41  W. Tao, A. Bur, J. L. Hockel, W. Kin, C. Tien-Kan, and G. P. Carman, Magn. Lett., IEEE **2**, 6000104 (2011).
42  P. Yang, K. Zhao, Y. Yin, J. G. Wan, and J. S. Zhu, Appl. Phys. Lett. **88**, 172903 (2006).
43  Y. Wang and Y. J. Jiang, Opt. Mater. **23**, 403 (2003).
44  A. W. Warner, J. Acoustical Soc. America **42**, 1223 (1967).
45  R. Zhang, B. Jiang, and W. Cao, J. Appl. Phys. **90**, 3471 (2001).
46  P. Han, W. Yan, J. Tian, X. Huang, and H. Pan, Appl. Phys. Lett. **86**, 052902 (2005).
47  J. Zhai, S. Dong, Z. Xing, J. Li, and D. Viehland, Appl. Phys. Lett. **89**, 083507 (2006).
48  D. Hasanyan, J. Gao, Y. Wang, R. Viswan, M. Li, Y. Shen, J. Li, and D. Viehland, J. Appl. Phys. **112**, 013908 (2012).
49  Y. Wang, D. Hasanyan, M. Li, J. Gao, J. Li, D. Viehland, and H. Luo, J. Appl. Phys. **111**, 124513 (2012).

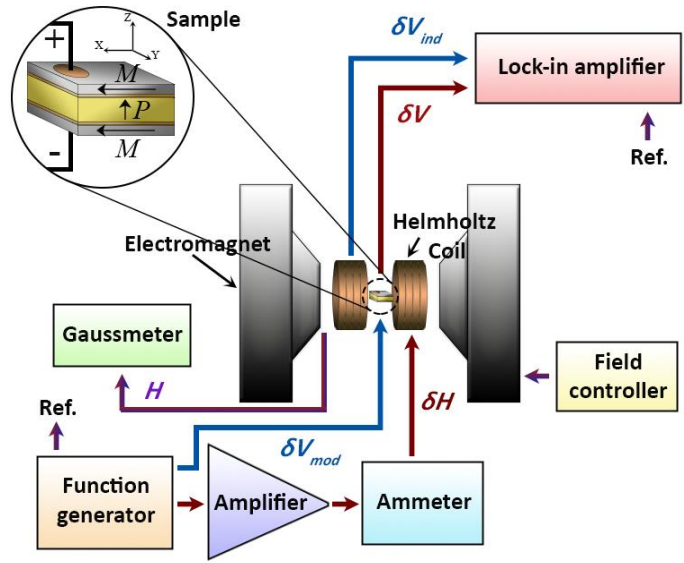

Fig. 1. (color online) Schematic presentation of the experimental setup. Red and blue arrows are associated with the measurements of the direct and converse ME effects, respectively.

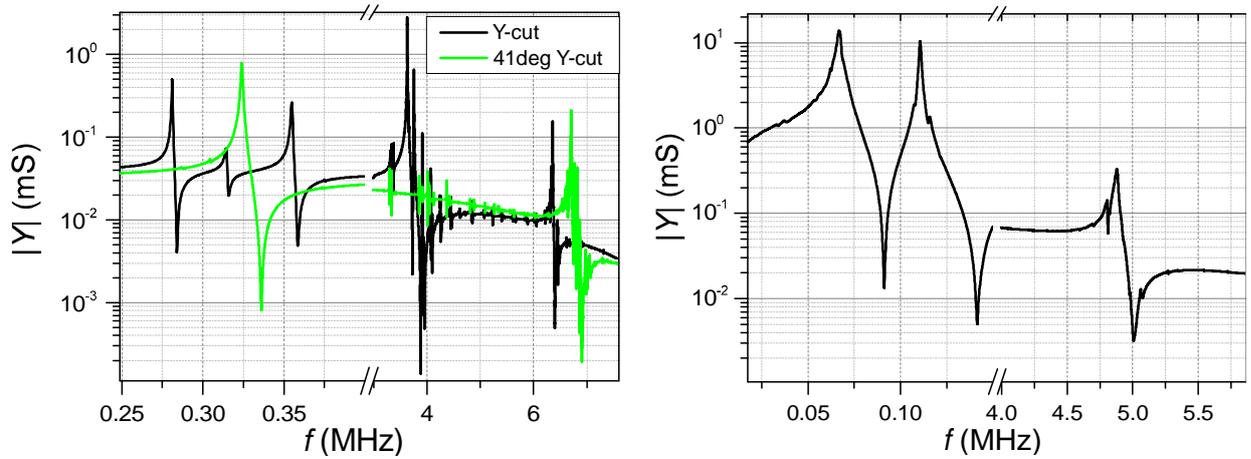

Fig. 2. (color online) Impedance spectroscopy of $LiNbO_3$ (left panel) and PMN-PT (right panel) based trilayers.

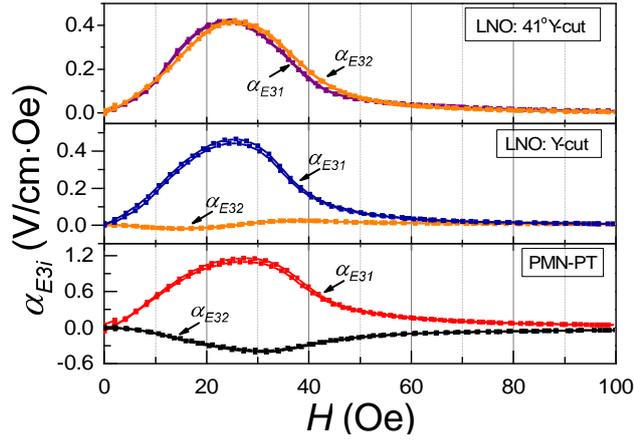

Fig. 3. (color online) Direct ME effect measurements at $f$ = 5 kHz. Each curve was measured under a bidirectional magnetic field sweep. Thus, a small hysteresis between up and down magnetic field sweeps can be seen on the measured curves.

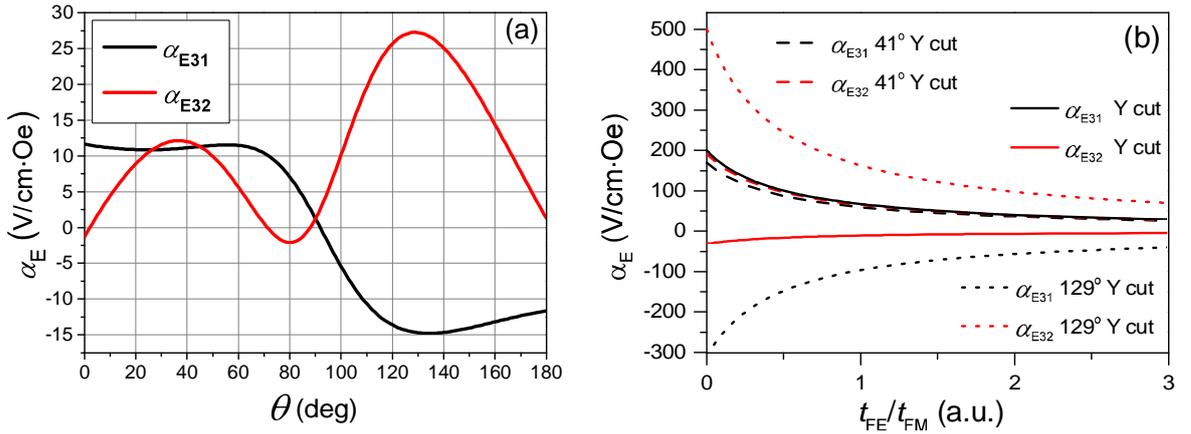

Fig. 4. (color online) (a) Calculated in-plane magnetolectric voltage coefficients as a function of the piezocrystal cut angle (read from the Y axis). (b) Magnetoelectric coefficients for different crystal cuts as a function of the ferroelectric/ferromagnetic relative thickness ratio.

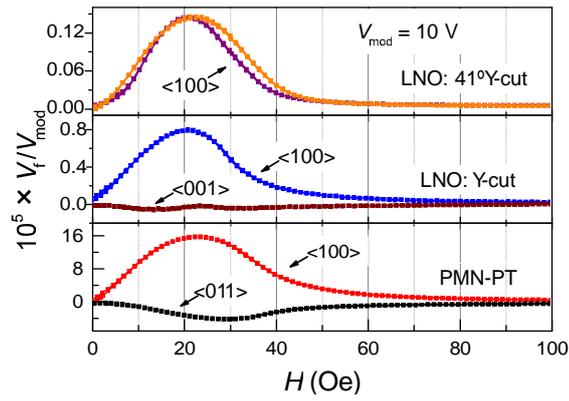

Fig. 5. (color online) Converse magnetoelectric effect (ratio of the voltage applied to the trilayer to the voltage generated on the sensing coil) measurements at $f$ = 5 kHz.

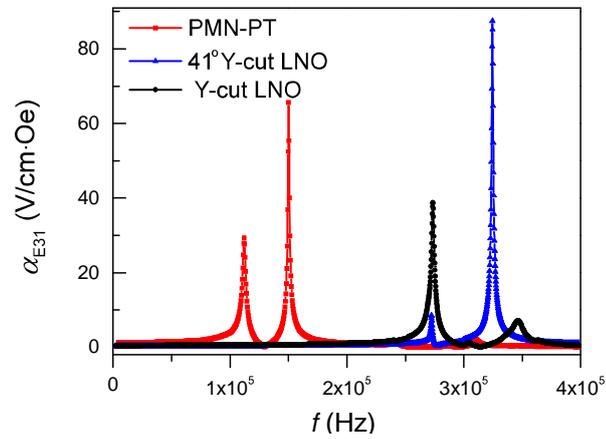

Fig. 6. (color online) Direct ME effect as a function of a frequency at $H$ = 30 Oe.